\documentstyle[12pt,aasms]{article}

\newcommand{\bq}{\begin{equation}}
\newcommand{\eq}{\end{equation}}

\begin{document}
 
\title{Time Dilation from Spectral Feature Age Measurements of  \\
        Type Ia Supernovae}

\author{A. G. Riess$^1$, A. V. Filippenko$^1$, D. C. Leonard$^1$, B. P. Schmidt$^2$, N. Suntzeff$^3$, M. M. Phillips$^3$, R. Schommer$^3$, A. Clocchiatti$^3$, R. P. Kirshner$^4$, P. Garnavich$^4$, P. Challis$^4$, B. Leibundgut$^5$, J. Spyromilio$^5$, R. C. Smith$^6$}
\small
\affil{$^1$Department of Astronomy, University of California, Berkeley, CA 94720-3411 \\
$^2$ Mt. Stromlo and Siding Springs Observatories, Australian National University, Private Bag, Weston Creek Post Office, ACT 2611, Australia \\ $^3$ Cerro Tololo Inter-American Observatory, Casilla 603, La Silla, Chile \\ $^4$  Harvard-Smithsonian Center for Astrophysics, 60 Garden Street, Cambridge, MA 02138\\ $^5$ European Southern Observatory, Karl-Schwarzschild-Strasse 2, D-85748 Garching, Germany \\ $^6$ University of Michigan, Department of Astronomy, 834 Dennison, Ann Arbor, MI 48109-1090}
\normalsize
\begin{abstract}
    We have developed a quantitative, empirical method for estimating the age of Type Ia supernovae (SNe Ia) from a single spectral epoch.  The technique examines the goodness of fit of spectral features as a function of the temporal evolution of a large database of SNe Ia spectral features.   When a SN Ia spectrum with good signal-to-noise ratio over the rest frame range 3800 to 6800 \AA $ \ $ is available, the precision of a spectral feature age (SFA) is (1-$\sigma$) $\sim$ 1.4 days.  SFA estimates are made for two spectral epochs of SN 1996bj (z=0.574) to measure the {\it rate} of aging at high redshift.  In the 10.05 days which elapsed between spectral observations, SN 1996bj aged 3.35 $\pm$ 3.2 days, consistent with the 6.38 days of aging expected in an expanding Universe and inconsistent with no time dilation at the 96.4\% confidence level.  The precision to which individual features constrain the supernova age has implications for the source of inhomogeneities among SNe Ia. 
\end{abstract}

\vfill
\eject
\section{Type Ia Supernovae as Clocks}

  A beguiling prediction of an expanding Universe is that distant objects will appear to age at a slower rate than nearby ones.  Type Ia supernovae (SNe Ia) provide extragalactic clocks of unparalleled precision which are sufficiently luminous to reveal this remarkable phenomenon.  While a few doubt that expansion alone causes cosmological redshifts (e.g., Arp 1987, 1994; Arp et al. 1990; Narlikar \& Arp 1993), the conventional interpretation has only modest experimental verification (Sandage \& Perelmuter 1991). 

Initial suggestions that time dilation might be seen in the photometric history of SNe Ia (Wilson 1939; Rust 1974) have been confirmed with recent observations of high redshift SN Ia light curves (Leibundgut et al. 1996; Goldhaber et al. 1997).  Specifically, Leibundgut et al. (1996) demonstrated that the broad light curve of SN Ia 1995K (z=0.479) was consistent with those of nearby SNe Ia
when dilated by a factor (1+z) as prescribed by cosmological expansion.

Yet recent work has shown that there is an intrinsic variation in the
breadth of SN Ia light curves which is related to the peak luminosity of
the supernova (Phillips 1993; Riess, Press, \& Kirshner 1995, 1996;
Hamuy et al. 1995, 1996a,b).  The sense of the correlation is that
more luminous SNe Ia give rise to broader light curves.  A skeptic
might postulate that within the large volume searched at high redshift,
we are naturally selecting the intrinsically brightest and broadest
light curves ever seen.  
 Although observations of the SN Ia
spectra can be used to eliminate such objects (Nugent et al. 1995; Filippenko et al. 1992),  Garnavich et al. (1996) and Maza et al. (1994) have seen that
atypical spectra do not always accompany peculiar light curves. 

Goldhaber et al. (1997) used five SN Ia light curves with 0.35 $\leq$ z
$\leq$ 0.46 to show that the {\it range} of observed light curve widths is
consistent with the sample of widths for nearby SNe Ia when dilated by (1+z).
For two of the objects they used an empirical relation between the $U-B$ color
and luminosity (Branch, Nugent, \& Fisher 1997) to identify the intrinsic width of
the light curve independently from the effect of time dilation.  Yet
the existence of such a relation is poorly determined due to the relative lack of ultraviolet data on SNe Ia, and it cannot guard against the existence of still
more luminous and broader SNe Ia than exist in the nearby sample.  A possible confusion between the effect of time dilation and luminosity variation is exacerbated by considering the latter as a ``stretching'' of the light curve.  

The temporal evolution of SN Ia spectra provides an alternate and
more reliable way to measure the passage of time for
individual SNe Ia.
After a detailed study of the nearby SN Ia 1937C, Minkowski (1939) noted that
``the spectra of different supernovae appear to differ so little, if
at all, that measurements of the wavelengths of the permanent bands
seem to provide a means of determining when a supernova reached its
maximum [brightness].''  Comprehensive multiwavelength studies of SN 1972E (Kirshner et al.
1973, Oke \& Searle 1974), SN 1981B (Branch et al. 1983), SN 1989B
(Barbon et al. 1990; Wells et al. 1994), and SN 1994D (Patat et al. 1996;
Meikle et al. 1996; Filippenko 1997) have confirmed what Minkowski had
surmised.  The temporal evolution of spectral features among most
SNe Ia proceeds at a remarkably reliable rate.  In addition, Branch, Fisher, \& Nugent (1993) have noted
that over 85\% of known SNe Ia display the typical features expected to
appear in ``normal'' SNe Ia.  This fraction increases to well over 90\% when subluminous SNe Ia (which are less likely to be found at large redshift) are eliminated.

  Following Minkowski (1939), we examine the idea that the pattern of features
in a single supernova spectrum can be used for an objective measure of a supernova's age.  Further, we suggest that two or more
spectra of the same supernova would be sufficient to measure the {\it rate}
at which an individual distant supernova ages as compared to that of
nearby SNe. Measuring the rate of aging with this method provides a direct way to
verify the presence of time dilation, expected to accompany an expanding
Universe.

   In \S 2 we determine the precision with which an SN Ia spectral
   feature age (SFA) can be
   measured with various spectral features.  We apply this technique
   in \S 3 to SN 1996bj, a SN Ia at z=0.574, determining its aging
   rate and the implications for time dilation. Section
   \S 4 discusses the results and their significance.

\section{Age Determination from SN Ia Spectra}

 Figure 1 shows the remarkable homogeneity of spectral features among the majority of SNe Ia when observed at the same age relative to $B$-band maximum. The uniformity of SNe Ia spectra, at a given phase, contrasts sharply with the spectral variations seen over a short interval of time.
     Figure 2 shows the rapid temporal evolution of some spectral features
     apparent in SNe Ia.  Certain
     portions of the SN Ia spectrum appear to change more quickly in
     time than
     others. Likewise, particular spectral features show a larger or smaller amount of
     variance from one supernova to another at a fixed age.  The
     spectral features which change rapidly with time and do not vary among
     different SNe Ia give the highest precision in determining the
     SFA.  Features whose shape and size differ among SNe Ia at a given
     age aid less in measuring the SFA but may give clues about the
     individual luminosity, temperature, and mass of the supernovae
     (e.g., Nugent et al. 1995; Fisher et al. 1995).  We discuss this further in \S 4.

    Here we explore the precision with which individual features and
    the spectrum as a whole constrain the age of the supernova.  We
    have chosen to treat the SN Ia spectrum as a number
    of short (200-400 \AA $ \ $wide) ``features'' to minimize the confusing effects of reddening, host galaxy contamination, and errors in flux calibration on the spectrum.  The features we use and their wavelength ranges are given in Table 1.  The feature boundaries were chosen to include the entire P-Cygni profile of assorted elements.

\begin{table}[h]    
\begin{center}    
\begin{tabular}{ccccccc} 
\multicolumn{7}{c}{Table 1: SN Ia Features} \\
\hline 
\multicolumn{4}{l}{} & \multicolumn{3}{c}{\em{S/N ratio}} \\
\multicolumn{5}{l}{} & \multicolumn{2}{c}{SN 1996bj} \\
  {\em Feature}     & range (\AA)  & elements  & $\sigma_{SFA}$ (days) & $\langle$ database $\rangle$  &  Oct 11 & Oct 21  \\

\hline   
 1 & 3800-4200 & Si II, Ca II & 3.07 & 22.7 & 22.7 & 12.0 \\
 2 & 4200-4580 & Mg II, Fe II & 2.45 & 35.7 & 20.8 & 12.2 \\
 3 & 4580-4950 & Fe II &  2.52 & 45.5 & 13.3 & 7.4 \\
 4 & 4950-5200 & Fe II & 2.14 & 55.6 & 8.3 & 6.6 \\
 5 & 5200-5600 & S II &  2.56 & 58.8 & 5.1 & 0.7 \\
 6 & 5600-5900 & Si II, Na I & 4.38 & 58.8 & na & na \\
 7 & 5900-6300 & Si II &  3.23 & 41.7 & na & na \\
 8 & 6300-6800 & Fe II & 2.74 & 45.5 & na & na \\
\hline 
\end{tabular}
\end{center}
\end{table}

\begin{table}[h]    
\begin{center}    
\begin{tabular}{ccl} 
\multicolumn{3}{c}{Table 2: Database of SNe Ia Spectra} \\
\hline 
SN Ia  & Julian Date of t$_{MAX}$(B) & Ages of Database Spectra\\
\hline   
SN 1981B & 2444671 & ($-$1)-1,17,20,24,29 \\
SN 1989B & 2447565 & ($-$7),($-$3)-($-$1),3,5,7-9,11-14,17-19 \\
SN 1990N & 2448083 & ($-$14),($-$7),7,14,17,38 \\
SN 1991M & 2448338 & -9 \\
SN 1992A & 2448641 & ($-$5),($-$1),3,5-7,9,11,16,17,24,28,37 \\
SN 1992G & 2448670 & 9,12,25 \\
SN 1994D & 2449431 & ($-$10)-($-$7),($-$5)-($-$2),2-6,8,11-14,16,18,20-23,25 \\
SN 1994M & 2449476 & 1,2,6,11,12,40 \\
SN 1994S & 2449519 & ($-$4),($-$3),1 \\
SN 1994T & 2449513 & 2 \\
SN 1994ae & 2449684 & 1-5,7-10,31,30 \\
SN 1995D & 2449769 & 3-5,7,9,11,14,16,32,38 \\
SN 1995E & 2449776 & ($-$4)-($-$2),0,5,8,31 \\
SN 1996X & 2450191 & ($-$3)-3,5-9 \\
SN 1996Z & 2450216 & 5 \\
SN 1996ai & 2450258 & ($-$4),($-$1) \\
\hline 
\end{tabular}
\end{center}
\end{table}

    We have compiled a database of high-quality
     spectra of all typical SNe Ia
     whose ages can be determined independently from their light curves. According to Branch, Fisher, \& Nugent (1993), ``typical'' SNe Ia spectra show conspicuous absorption features near 6150 \AA $\ $due to Si II and
near 3750 \AA $\ $ from Ca II near maximum light.  These features are missing or notably weak among abnormal SNe Ia spectra.  For the over 85\% of SNe Ia which are typical, we expect the change in spectral features to correspond with a singular progression in age.  For this analysis, we exclude the small number of ``atypical'' SNe Ia whose spectral features are likely to vary both with age and intrinsic SN Ia charactersistics in a complex way. 

In Table 2 we list the 126 spectra which comprise our database of SN Ia spectra.  We can determine the age of each spectrum from the time since $B$-band maximum light.  Column 2 lists the Julian dates of $B$-band maximum as determined by the MLCS fit to the light curves (Riess, Press, \& Kirshner 1996).  The uncertainty in the date of maximum is $\sim$ 1 day. Column 3 gives the ages of SN Ia spectra comprising the database of spectra.  The spectra in the database
     sample the range in age of fourteen days before maximum to thirty-eight days after maximum.  After forty days,
     changes in SN Ia spectra occur more gradually, making it
     difficult to make a precise determination of SN Ia ages from spectra.
The mean signal-to-noise (S/N) ratios of the spectra are listed in column 5 of Table 1.  Each of the
     spectra has been shifted to zero velocity using the measured redshift
     of the host galaxy. 
  
   To estimate a spectrum's unknown SFA, our procedure is to
   measure its features' goodness of fit
   to all the spectral database of features with known age.  The goodness
 of fit is the minimum $\chi^2$ resulting from varying a feature's mean
   flux to match the mean flux of the feature in the database.  In
   addition to accounting for differences in SN Ia apparent flux, this
   procedure guards against the effects of errors in spectral flux
   calibration.  Indeed, it permits one to estimate the age of a supernova in a short amount of time, while observing at the telescope.  The
   noise in a spectral feature is
   given by the variance between a feature and the same feature
   smoothed over a 30 \AA $\ $bin (a size sufficiently small to
   leave the spectral undulations unaffected).  Undated spectra are not fit to the lowest S/N (i.e., $\leq$ 10) members of the database whose noise estimates are too uncertain.

   A simple procedure is utilized to locate the age at which the $\chi^2$ function of age is a minimum,  and a number of precautions are exercised to insure that the inferred
   age is reliable.  The minimum is given by the $\chi^2$-weighted average of the 4 ages with lowest $\chi^2$.  This minimum must not be significantly higher than its expectation value.  This requires that at some age the undated feature resembles
   objects in the database.  With this criterion, we cannot determine the SFA for spectroscopically
   peculiar SNe Ia or objects which are not SNe Ia.  The value of the $\chi^2$ minimum must rise substantially (in this case, 20\%) within
   five days of the inferred age.  This caution excludes
   features whose minima are too shallow to give a robust
   SFA.  A corollary to the above is to reject minima which occur on or near the
   extremes of the age range represented in the database.  In
   adherence to this requirement, we only accept minima located
   between seven days prior to maximum and twenty-five days after
   maximum.  Despite the rigor of these precautions, we can measure a reliable SFA for over 80\% of the spectra in the database.

   We can test the precision of SFAs by removing each of the spectra
   from the database and comparing its measured SFA to its light curve
   age.  The results are very encouraging.  We have
   quantified the dispersion between the age measured from the light curve and the SFA determined from each individual
   feature and listed them in column 4 of Table 1.  The dispersion for the SFA derived from a single feature ranges from 4.4 days for ``feature 6'' centered on the Si II absorption at 5850
   \AA $\ $ to a low of 2.1 days for ``feature 4'' encompassing some iron-peak elements.  The age uncertainty derived from the assorted spectral features is shown in Figure 3 with an example of those features.  

   The best use of the spectrum for deriving the supernova age is to combine all available features to measure the age.  To this end, we estimated the age where the $\chi^2$ fit to all of the useful features was minimized.  For this purpose we used all features which were observed in the spectra and whose individual $\chi^2$ minimum was not located at the extreme of the range of ages available in the database.     

   The SFAs derived from the entire spectrum are shown in Figure 4.  The dispersion between the light curve ages and the full spectrum SFA is 1.70 days.  This dispersion comes from the variance of both the SFA measurement and the light curve age determination.  The uncertainty in the ages derived from the light curves is $\sim$ 1.0 day, making the {\it true} SFA uncertainty (1-$\sigma$) $\sim$ 1.4 days. 

   We can estimate the uncertainty of the SFA independently from the uncertainty in the light curve age by comparing the two age estimates separately for each SN Ia.  Such a comparison is shown in Figure 5 for SN 1989B.  The error made in estimating the light curve age of a given SN Ia is removed by fitting an offset between the ages.    The remaining dispersion is solely from the SFA estimate.  This yields a true SFA uncertainty (1-$\sigma$) of $\pm$ 1.3 days for the spectra in the database for which we have multiple epochs, and confirms the $\sim$ 1.0 day uncertainty in the light curve age.  We will conservatively adopt $\sigma_{SFA}$=1.4 days.  

 One might assume that eight independent age estimates with uncertainties listed in Table 1 (column 4) would, when combined, yield a precision of better than  1 day.  This is not the case because the uncertainties in the SFA derived from the individual features are strongly correlated.  There are a number of causes for correlations.  First, because the SFA from each feature is compared to the same imprecise light curve age the differences are related.  Second, it seems likely that the idiosyncracies of individual SNe Ia cause the features to systematically overpredict or underpredict the age in similar ways.  Finally, not all eight features are necessarily available in every spectrum, somewhat diminishing the quality of the SFA estimate.

Nevertheless, the precision of an SFA is remarkable.  Given the spectrum of a SN Ia of similar quality as those in the database, we can make a prediction of the supernova's age with an uncertainty of only 1.4 days.  Although in principle one could also derive the redshift from a SN Ia spectrum, in practice the recession velocity is best measured from the narrow emission and absorption features of the host galaxy spectrum.

    SNe Ia are tools with great leverage for measuring distances of cosmological significance. Distance uncertainties range from 5\% to 10\% when the light and color curves are used to account for variations in luminosity and extinction (Phillips 1993; Riess, Press, \& Kirshner 1995, 1996; Hamuy et al. 1995, 1996a).  Yet, to extract the information needed from the light curves to measure a precise distance, photometric observations must begin within 10 days of maximum light (Riess, Press, \& Kirshner 1996).  For this reason, a precise SFA from the very first spectrum of each SN would be an invaluable tool to determine which SNe Ia are worth following with one's limited observing resources.  Further, SFA estimates made from any number of a given supernova's spectra could tighten constraints on the date of maximum used in fitting its light curve shape.

Recently, two groups have embarked on ambitious programs to detect and measure the distances to SNe Ia at redshifts of 0.3 $\leq$ z $\leq$ 0.8 (Schmidt 1997; Perlmutter et al. 1997).   With SFAs of two or more spectra, we can hope to measure the {\it rate} of aging.  For SNe Ia in an expanding universe, we expect to find the rate of aging {\it reduced} by the factor (1+z).

\section {Time Dilation Detected by Spectral Feature Aging}

  SN 1996bj was discovered by the High-Z Supernova Search Team (IAUC 6490).  A spectrum obtained on 1996 October 11 at UT 11:50 at the Keck II telescope with the Low Resolution Imaging Spectrometer (LRIS; Oke et al. 1995) by A. V. Filippenko, D. C. Leonard, A. J. Barth, and C. Y. Peng showed SN 1996bj to be a Type Ia supernova at z=0.574 (as determined by narrow emission from the host galaxy).  A second spectrum of SN 1996bj was obtained on October 21 at UT 12:55 at the Keck II telescope by A. V. Filippenko. D. C. Leonard, A. G. Riess, and S. D. Van Dyk.  Both spectra are shown in Figure 6, shifted to zero velocity.  Due to the high redshift of SN 1996bj, only the rest wavelength range 3300 to 5600 \AA $ \ $ was observable.  Although the Si II absorption feature at 6150 \AA $ \ $ is beyond this range, the visible characteristics of the spectrum show that SN 1996bj is atleast moderately typical of SN Ia.  The presence of Ca II absorption ($\sim$ 3750 \AA $ \ $) and the absence of obvious Ti II absorption ($\sim$ 4200 \AA $ \ $) indicates that SN 1996bj is neither like the atypical and overluminous SN 1991T (Filippenko et al 1992a, Phillips et al 1992) nor like the unusual underluminous SN 1991bg (Filippenko et al 1992b, Leibundgut et al 1993).  In an expanding universe, one would expect SN 1996bj to have aged only 6.38 days during the 10.04 days that elapsed between successive spectra.  With the use of the SFAs we can test this prediction.

   We have used the procedure described in $\S$ 2 to measure the SFAs of the two spectra of SN 1996bj.  Due to the extreme redshift of the supernova, only Features 1 through 5 occur at wavelengths where the spectrometer is sensitive .  The SFA measured from these features in the October 11 spectrum is 5.55 days before maximum.  The same five features in the October 21th spectrum yield a SFA of 2.20 days before maximum. The goodness of fit as a function of the SFA for each spectrum is shown in Figure 7.  Figure 6 depicts examples of features from the spectral database which fit well to the features in the spectra of SN 1996bj.

What is the uncertainty of these measurements?  If the two spectra of SN 1996bj had the wavelength coverage to include all eight features (in this case 6000 \AA $\ $ to 10700 \AA) at a S/N ratio comparable to the spectra comprising the database, the SFA uncertainties would be $\pm$ 1.4 days, as determined in \S 2.  Restricting the database spectra to the same first five features results in an increased SFA uncertainty of $\pm$ 1.6 days.  Next, we add noise to each of the database spectra to match the S/N ratio of the SN 1996bj spectra (Table 1, columns 6 \& 7) and measure their SFAs.  Database spectra which include the same five features at the same S/N ratio as the Oct 11 and 21 spectra of SN 1996bj yield SFA uncertainties of $\pm$ 2.1 and $\pm$ 2.4 days, respectively.    This is the uncertainty we expect for the SFA measurements of SN 1996bj.

 The difference between the two SFA ages of SN 1996bj is 3.35 $\pm$ 3.2 days.  This is the measured amount of time which has appeared to have elapsed in the supernova's life.  It is significantly briefer than the 10.05 days which passed between successive spectral observations.  The {\it reduced} aging for SN 1996bj as compared to nearby SNe Ia is consistent with the 6.38 days of aging expected in an expanding Universe and is inconsistent with the null hypothesis (i.e., no time dilation) at the 96.4\% confidence level.  

\section{Discussion}

  Individual features vary significantly in their ability to constrain the age of the supernova.  These differences offer important clues about the source of inhomogeneities among SNe Ia.  For a feature to be a good indicator of SN Ia age, it must change quickly with time and its characteristics must not vary among different supernovae.  Features which give imprecise SFAs, yet evolve quickly with time, should provide the best means to discern intrinsic variations among SNe Ia.  Two of the latter features, numbers 6 and 7, are the Si II absorption troughs at 5800 \AA $\ $ and 6150 \AA $\ $ which give SFA uncertainties of 4.4 and 3.2 days, respectively.  Nugent et al. (1995) demonstrated that the ratio of these two features provides a useful measure of intrinsic SN Ia luminosity.   Fisher et al. (1995) found a similar relation between luminosity and the Ca II H \& K absorption which is located in Feature 1, another imprecise feature for SFA. Our findings support these results; in fact, our analysis is similar to a demonstration by linear regression that a multivariate analysis is needed.  In our case, the imprecision of individual feature SFAs reveals the presence of intrinsic SN Ia variables at work.  Given recent evidence that SN Ia luminosities have a significant internal dispersion, one might infer that features which give imprecise SFAs could be useful indicators of SN Ia luminosity.

Features 2, 3, 4, and 5 give the best measures of SFA with respective uncertainties of only 2.5, 2.5, 2.1, and 2.6  days (Fig. 2). These features are narrower and arise from lower velocity ejecta than the features which give less precise SFAs.
  Further, the better predictors of SFA are dominated by Fe II while the poor SFA discriminators are primarily from Si II absorption.  We conclude from this that near the iron core SNe Ia are quite similar, but differences arise closer to the surface where the intermediate-mass elements are ejected at high velocity .  This description fits well with the observations that the intermediate-mass element production which occurs near the surface varies within the SN Ia class (Filippenko et al. 1992a,b; Leibundgut et al. 1991, 1993; Phillips et al. 1992).  It is also consistent with the observation that late-time nebular phase spectra of SNe Ia which probe deep into the iron core are remarkably homogeneous (Ruiz-Lapuente et al. 1992)  Still, it is likely that small intrinsic differences among SNe Ia arise even in the most homogeneous of features, negating the possibility of making a nearly perfect SFA estimate.

In the future, it should be possible to improve SFA estimates by learning to discriminate between the effects of age and explosion characteristics on the SN Ia spectral features. A growing number of atypical SNe Ia, exemplified by SN 1991bg and SN 1991T, may provide clues to the role that temperature, luminosity, and abundance differences play in shaping the spectral features of SNe Ia.   All supernovae in our current analysis show the characteristic features seen in typical SNe Ia spectra and span a moderate range of decline rates (0.9 $\leq$ $\Delta$M$_{15}$$(B)$ $\leq$ 1.6).  With the present analysis, we can use the spectral features seen in typical SNe Ia to measure a precise SFA.  For such SNe Ia, we see no relation between a supernova's light curve shape and any error made in predicting its SFA.  With the current database of available SNe Ia, we recommend against measuring a SFA for atypical SNe Ia such as SN 1991T and SN 1991bg.  Our experience is that if a SFA can be measured for such objects, it is often flawed and should not be trusted.

The uncertainty we assign to any SFA measurement depends only on the S/N ratio of the spectrum.  It is determined from the dispersion of SFAs for database spectra with similar signal-to-noise ratios.  On the average this method properly quantifies the uncertainty in an SFA.  Unfortunately, the standard practice of measuring the SFA uncertainty from the variation of $\chi^2$ with age is not appropriate.  The database spectra which we fit to an unknown spectrum are not a function of age, they are merely {\it labeled} by age.   Our model is not a one-parameter family, but rather a set of discrete models which provide a set of goodness-of-fits.  Although the curvature of the $\chi^2$ is suggestive of the uncertainty in age, it cannot directly quantify it.

One more snapshot of the aging process of SN 1996bj is, in principle,  available to us.  All spectral feature ages have been defined by us relative to the photometric maximum of the light curve.  ``Maximum'' is an event whose spectral and photometric definitions coincide.  Therefore, locating the time of the photometric maximum is equivalent to obtaining a third spectral epoch of SN 1996bj at maximum.  The use of the maximum along with the two SFAs for determining the rate of aging of SN 1996bj remains completely independent of the light-curve-width time dilation test of Leibundgut et al. (1996) and Goldhaber et al. (1997).  Unfortunately, the time of maximum light of SN 1996bj is poorly constrained by the photometric data available to us, and for this case it offers no help in measuring the rate of aging.  We can only say that the time of maximum predicted by the two SFAs is consistent with the range of possible times of maximum
of the light curve.

How much of an improvement does our rigorous treatment of spectral feature aging provide over the ``old-fashioned'' method of estimating by eye?  While it is hard to quantify the precision of a visual estimate, a recent anecdote is illuminating.   IAU Circular 6381 contained five independent identifications of SN Ia 1996X made from spectra obtained on the same night, 1996 April 14.  Two teams said the supernova was ``near maximum,'' one said ``before maximum,'' another thought one week before maximum, while the last said two days after maximum.  The standard deviation of these estimates appears to be larger than the 1.4 days we can attain with a quantitative SFA, although some improvement in visual age estimates could certainly be made in the future through careful scutiny of SN Ia spectra.  For comparison, the light curve age of SN 1996X on the same day was 3 $\pm 1$ day before maximum while the SFA age was 1.3 $\pm 1.4$ days before maximum.  

Besides measuring time dilation at large redshifts, SFAs can provide a way to determine the progress of a SN Ia along its light curve.  This information is useful to astronomers observing SNe Ia to measure cosmological distances.  With refinement of this technique, it might become possible to reduce the amount of observing necessary to measure the distance to a SN Ia; in principle, a single spectrum and one or more photometric measurements might suffice.  With this strategy, certain spectral features could constrain the supernova age while others could determine the luminosity. A few photometric epochs would complete the information necessary to measure the distance.  If a multiply lensed SN Ia is ever observed, SFA would provide a very precise measure of the time delay between the light paths of the images.

 Finally, there are three natural ways to sharpen this direct probe of time dilation.  One is to include more spectral epochs of observation of the high-redshift SNe Ia.  For SN 1996bj, inclement weather foiled our attempts to do this.  Equivalently, one can add more SNe Ia to this analysis; at the current rate of discovery of high-redshift SNe Ia, this is likely to occur.  Moreover, improving the S/N ratio with longer integrations, as well as including the three features at longer wavelengths, could increase the precision of an SFA to the 1.4 day precision attainable for nearby SNe Ia.  With improvements of the nature outlined above, spectral feature age measurements of Type Ia supernovae will provide the means to strongly constrain the rate of aging for objects at high redshift.  
\\
\\
The W. M. Keck Observatory, made possible by
the generous and visionary gift of the W. M. Keck Foundation, is operated
as a scientific partnership between the University of California and
the California Institute of Technology. We are grateful to the Keck staff, as well as A. J. Barth, C. Y. Peng, and S. D. Van Dyk, for their assistance
with the observations. This work was supported by the NSF through
grant AST--9417213 to A.V.F., and by the Miller Institute for Basic Research in Science through a fellowship to A.G.R.  We would like to acknowledge valuable discussions with Peter Nugent and Alex Kim.  
\end{document}